\documentclass[11pt]{article}
\usepackage{amsmath,amssymb}
\usepackage[pdftex]{graphicx}
\usepackage[pdftex]{color}
\usepackage[top=30truemm,bottom=30truemm,left=25truemm,right=25truemm]{geometry}

 \def\ep{{\epsilon}}

 \def\frac#1#2{{#1\over #2}}

\def\be{\begin{equation}}
\def\ee{\end{equation}}
\def\ba{\begin{eqnarray}}
\def\ea{\end{eqnarray}}

 \def\f {\frac}
 \def\ti{\tilde}

 \def\no{\nonumber \\}

 \def\ep{\epsilon}

\def\mat{\mathcal}

\usepackage{color}
\usepackage{graphicx}
\usepackage{dcolumn}
\usepackage{bm}

\begin{document}
\thispagestyle{empty}

\begin{flushright}
YITP-18-68\\
\end{flushright}

\vspace{.4cm}
\begin{center}
\noindent{\Large \textbf{Time Evolution after Double Trace Deformation
}}\\
\vspace{2cm}

Masamichi Miyaji\vspace{1cm}

{\it
Yukawa Institute for Theoretical Physics (YITP),
Kyoto University, \\
Kitashirakawa Oiwakecho, Sakyo-ku, Kyoto 606-8502, Japan
}

\vskip 2em
\end{center}

\vspace{.5cm}

\begin{abstract}
In this paper, we consider double trace deformation to single CFT${}_2$, and study time evolution after the deformation.
The double trace deformation we consider is nonlocal: composed of two local operators placed at separate points.
We study two types of local operators: one is usual local operator in CFT, and the other is HKLL bulk local operator, which is still operator in CFT but has properties as bulk local operator.
We compute null energy and averaged null energy in the bulk in both types of deformations. 
We confirmed that, with the suitable choice of couplings, averaged null energies are negative. This implies causal structure is modified in the bulk, from classical background.
We then calculate time evolution of entanglement entropy and entanglement Renyi entropy after double trace deformation.
We find both quantities are found to show peculiar shockwave-like time evolution.

\end{abstract}

\section{Introduction and Summary}

 Recent studies revealed that, introduction of non local double trace deformation can render non traversable wormholes traversable\cite{Gao:2016bin}. Wormholes considered in these studies are two sided black hole connecting two asymptotically AdS regions. In AdS/CFT the geometry is dual to thermofield double state on ${\rm CFT}_L\otimes {\rm CFT}_R$. The non local double trace deformation added to the CFT action is
\be
g\int dt~d^dx~h(t,x)\mat{O}_{L}(t,x)\mat{O}_{R}(-t,x),
\ee
where operator $\mat{O}_{L/R}$ lives in $CFT_{L/R}$. Such deformation causes null shockwave in the bulk, whose null energy can be positive or negative according to the sign of coupling $g$. When the averaged null energy is negative, it was found that the shock wave opens up traversable wormhole in non traversable wormhole, even if $g$ is infinitesimally small. In this sense, wormholes in two sided black hole are barely non-traversable. Applications and generalizations of these double trace deformations result in interpretation as quantum teleportation\cite{Maldacena:2017axo}, exploration into black hole interior\cite{Almheiri:2018ijj}\cite{deBoer:2018ibj}, and realization of eternal wormhole\cite{Maldacena:2018lmt}.
 
 Traversability of wormhole is intimately related with violation of averaged null energy condition (ANEC).
 Averaged null energy condition is a conjecture, that states integral of null energy on null ray must be non-negative in any UV complete QFT. ANEC is proven on Minkowski spacetime and other special cases\cite{Kelly:2014mra}\cite{Faulkner:2016mzt}\cite{Hartman:2016lgu} and its generalization QNEC\cite{Bousso:2015wca}\cite{Koeller:2015qmn}\cite{Wall:2017blw}\cite{Balakrishnan:2017bjg}. Averaged null energy has simple intuitive interpretation, that it can measure the change of causal structure, when we perturb solution of vacuum Einstein equation by matter stress tensor.
 When ANE is negative, null ray in unperturbed metric becomes timelike in perturbed metric. Related to this fact, it is known that in classical general relativity, existence of traversable wormhole implies negative ANE. The results of \cite{Gao:2016bin}\cite{Maldacena:2017axo} are consistent with ANEC of QFT, since once one applies double trace deformation to the theory, the theory is no longer QFT nor Lorentz invariant, although it is unitary .
  
 Interesting generalization of previous studies on double trace deformation is to consider deformation composed of HKLL bulk local operators\cite{Hamilton:2006az}, not the local operators of boundary CFT. With this deformation, one can manipulate AdS geometry directly, not indirectly through AdS boundary. Stating differently, one can create bulk local objects which are connected by quantum entanglement. Double trace deformation can be obtained effectively when we integrate out fields which are not relevant to the energy scale under consideration\cite{Maldacena2018}. Study on bulk local double trace deformation may help to understand realizing traversable wormhole in pure AdS, or to understand non local effects of higher derivative interactions in string theory.
 
In order to understand double trace deformations in more detail, it is also important and interesting to study time evolution of field theory observables in general. In particular, it is interesting to understand the dynamics of entanglement entropy and entanglement Renyi entropy, with non local double trace deformation.
  
 In this paper, we study time evolution of bulk AdS and boundary CFT, after double trace deformation. We focus on double trace deformation composed of boundary local operators, as well as HKLL bulk local operators. In both cases, we confirm that with appropriate choice of sign of $g$, we obtain negative averaged null energy. This implies time advance or time shifts, meaning originally light-like separated points become time like after incorporating the effects of the matter. Further, we confirm that HKLL bulk local double trace deformation can act as point like source of negative null energy. 
 We then study time evolution of general CFT${}_2$ after double trace deformation, by computing leading correction to entanglement entropy and entanglement Renyi entropy.
 We found the leading correction is universal, in other words it depends only on conformal dimension of double trace deformation. Their time dependence show interesting behavior will be described.
  
  In section 2, we consider pure AdS and introduce double trace deformation which connects antipodal points in the boundary CFT. We will follow the time evolution after the quench by the double trace deformation, focusing on null energy in the bulk. We confirmed that averaged null energy on null geodesic is linear in $g$ so in particular it can be negative.
  
 In section 3, we consider new type of double trace deformation, composed by HKLL bulk local operators.
 This deformation introduces direct connection between two distinct bulk points. The motivation is to initiate the study of traversable wormholes in spacetimes with single asymptotic region using AdS/CFT.
 We then study the spacetime structure after the deformation, in particular null energy in the bulk. We found that, as in the case of section 2, averaged null energy on null geodesic is linear in $g$ so in particular it can be negative.
  
 In section 4, we consider time evolution of entanglement entropy and entanglement Renyi entropy after double trace deformation. To calculate entanglement entropy, we use known expression of modular Hamiltonian. And for Renyi entropy, we use Replica trick. In both cases, the leading corrections from the double trace deformation to vacuum are proportional to $g$. This means one can increase or decrease the entanglement of subsystems by double trace deformation.
 
 In the final section, we explain implication of our results and future directions. 
 

\section{Double trace deformation of boundary local operators}
In this section, we consider the time evolution of matter in bulk after introducing double trace deformation to boundary CFT. We consider holographic CFT dual to classical gravity on pure global AdS${}_3$
\be
ds^2=\f{1}{cos^2\rho}(-d\tau^2+d\rho^2+sin^2\rho d\theta^2),
\ee
where $0\leq \rho \leq \f{\pi}{2}$, $-\pi<\theta<\pi$ and $-\infty<\tau<\infty$.\\
We deform CFT Hamiltonian by double trace deformation
\be
H\rightarrow H^g(t):=H+g\mat{O}(0,0)\mat{O}(0,\pi)f(t),
\ee
 which is designed to connect two distinct points $\theta=0$ and $\theta=\pi$ at AdS boundary.
 $\mat{O}$ is scalar primary operator of the CFT with conformal dimension $\Delta=\delta+\bar{\delta}$.
 $g$ is a deformation parameter and infinitesimally small, and $f(t)$ is some real function that only depends on $t$.\\

Such double trace deformation does not produce non zero scalar fields at the leading order in $g$, but does produce 
nontrivial configuration of matter stress tensor in the bulk, which we will study below.
From now on, we assume large central charge, so that the bulk gravity is classical Einstein gravity.
We also ignore the back reaction to gravity from matter fields. 
However, we will treat bulk matter quantum mechanically.
 
 We first compute perturbed bulk two point function, using HKLL bulk local operator.
Two point function of HKLL bulk local operators satisfies
 \be
 (\nabla^2_{AdS_3}-M^2)\langle\Phi(t,\rho,\theta)\Phi(t',\rho',\theta')\rangle=\f{1}{\sqrt{-g^{(3)}}}\delta^3((t,\rho,\theta),(t',\rho',\theta'))
  \ee
  with boundary condition
  \be
  \f{1}{(cos\rho)^{\Delta}}\Phi(t,\rho,\theta) \underset{\rho\rightarrow \f{\pi}{2}}{\rightarrow}
  \mat{O}(t,\theta).
  \ee
where $M^2=\Delta(\Delta-2)$.
 
 Let us first consider time evolution of HKLL bulk local operator.
 After deformation, it is given by
 \ba
 \Phi^{g}(t,\rho,\theta)&&=\ti{P}e^{i\int^{t}_{-\infty}d{\ti t}~ e^{iH\ti{t}}(H^g(\ti{t})-H)e^{-iH\ti{t}}} \Phi(t,\rho,\theta) Pe^{-i\int^{t}_{-\infty}d{\ti t}~e^{iH\ti{t}}(H^g(\ti{t})-H)e^{-iH\ti{t}}}\no&&
 = \Phi(t,\rho,\theta)+i\int^t_{-\infty}d\ti{t}~[e^{iH\ti{t}}(H^g(\ti{t})-H)e^{-iH\ti{t}},~ \Phi(t,\rho,\theta)]+\mat{O}(g^2). \ea
 Then the first order change of bulk two point function is given by
\ba
&&\langle\Phi^g(t,\rho,\theta)\Phi^g(t',\rho',\theta')\rangle-\langle\Phi(t,\rho,\theta)\Phi(t',\rho',\theta')\rangle\no&&=ig\int^t_{-\infty} d\ti{t}~f(\ti{t})\langle [\mat{O}(\ti{t},0)\mat{O}(\ti{t},\pi),\Phi(t,\rho,\theta)]\Phi(t',\rho',\theta')\rangle+c.c_{(t,\rho,\theta)\leftrightarrow(t',\rho',\theta')}+\mat{O}(g^2)
\no&&\approx
ig\int^t_{-\infty} d\ti{t}~f(\ti{t})\Big(\langle [\mat{O}(\ti{t},0),\Phi(t,\rho,\theta)]\rangle\langle\mat{O}(\ti{t},\pi)\Phi(t',\rho',\theta')\rangle
\no&&
+\langle [\mat{O}(\ti{t},\pi),\Phi(t,\rho,\theta)]\rangle\langle\mat{O}(\ti{t},0)\Phi(t',\rho',\theta')\rangle\Big)
+c.c_{(t,\rho,\theta)\leftrightarrow(t',\rho',\theta')}+\mat{O}(g^2)
\ea
In the final line we used large N approximation.

Once we know the bulk two point function, we can compute expectation value of bulk stress tensor perturbatively.
 The bulk energy momentum tensor is
 \be
 T_{\mu\nu}=\partial_{\mu}\phi\partial_{\nu}\phi-\f{1}{2}g_{\mu\nu}g^{\rho\sigma}\partial_{\rho}\phi\partial_{\sigma}\phi-\f{1}{2}g_{\mu\nu}M^2\phi^2,
 \ee
so the expectation value of stress tensor can be obtained by point splitting method with 
coincident point divergence subtracted, as follows,
\be
\langle T_{\mu\nu}\rangle=\underset{x\rightarrow x'}{\rm Lim} \Big(\partial_{\mu}\partial_{\nu}'G(x,x')-\f{1}{2}g_{\mu\nu}g^{\rho\sigma}\partial_{\rho}\partial_{\sigma}'G(x,x')-\f{1}{2}g_{\mu\nu}M^2G(x,x')\Big),
\ee
where $G(x,x'):=\langle\Phi(x)\Phi(x')\rangle$.

Let us assume $f(t)=\delta(t)$, so that the double trace deformation is instantaneous. Then we can obtain analytical expression for bulk stress tensors,
using known expression of bulk two point function (see appendix A).
In order to explore dynamics of bulk matter, we will focus on $\langle T_{tt}\rangle$ and null energy in the bulk, integral of $\langle T_{tt}\rangle$ on Cauchy slices, and integral of null energy on null geodesics.
 
 Averaged null energy is defined as integral of null energy on complete null geodesic.
We will consider null geodesics passing through the center of global AdS$_3$, starting at $t=t_0$ and $\theta=\theta_0+\pi$, ending at $t=t_0+\pi$ and $\theta=\theta_0$.
\begin{figure}  
                  \begin{center}
		         \includegraphics[scale=1]{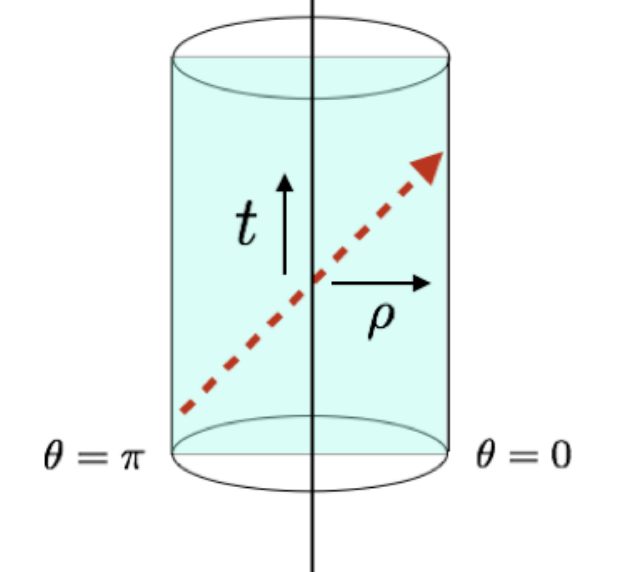} 
		         \label{1}
		         		        		        	 \end{center} \label{coordinate}
			  \caption{The geometry of global AdS${}_3$. Red line is null ray, on which we compute averaged null energy. On blue plane we compute null energy, and null direction is set as that of the red line.}
			\end{figure}

\begin{figure}  
                  \begin{center}
		         \includegraphics[scale=0.9]{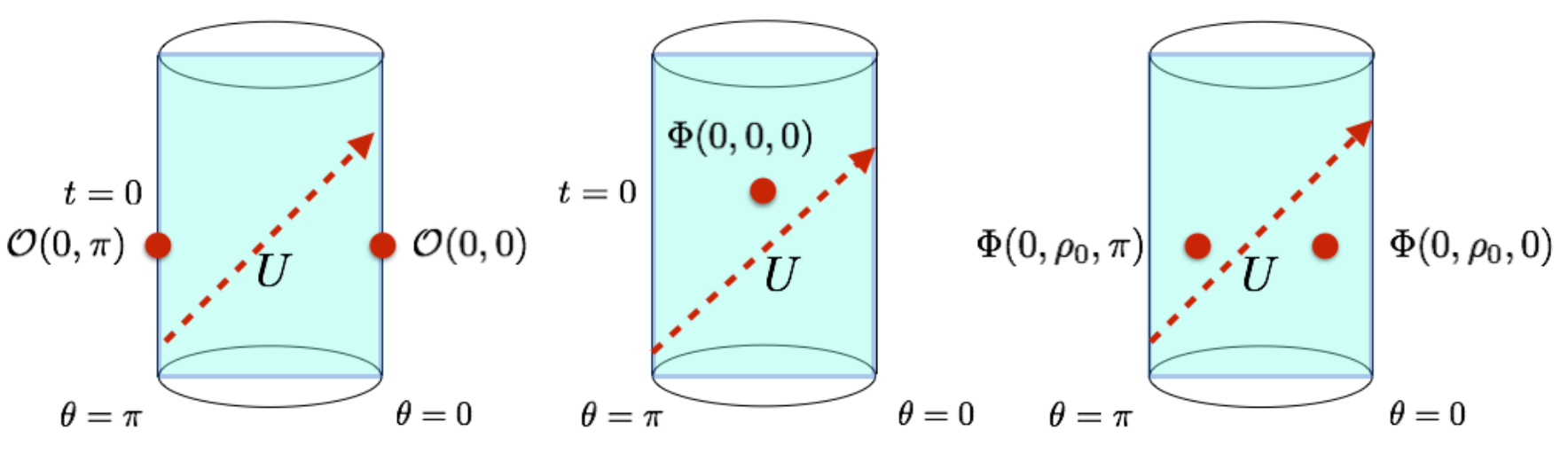} 
		        		        	 \end{center} 
					 \label{2}
			  \caption{The configurations of double trace deformations we consider in this paper. (Left) Double trace deformation composed of boundary local operator.(Center) Single trace deformation in the bulk. (Right) Double trace deformation pushed into the bulk.}
			\end{figure}
			     
\begin{figure}     
			                     \begin{center}
		         \includegraphics[scale=0.7]{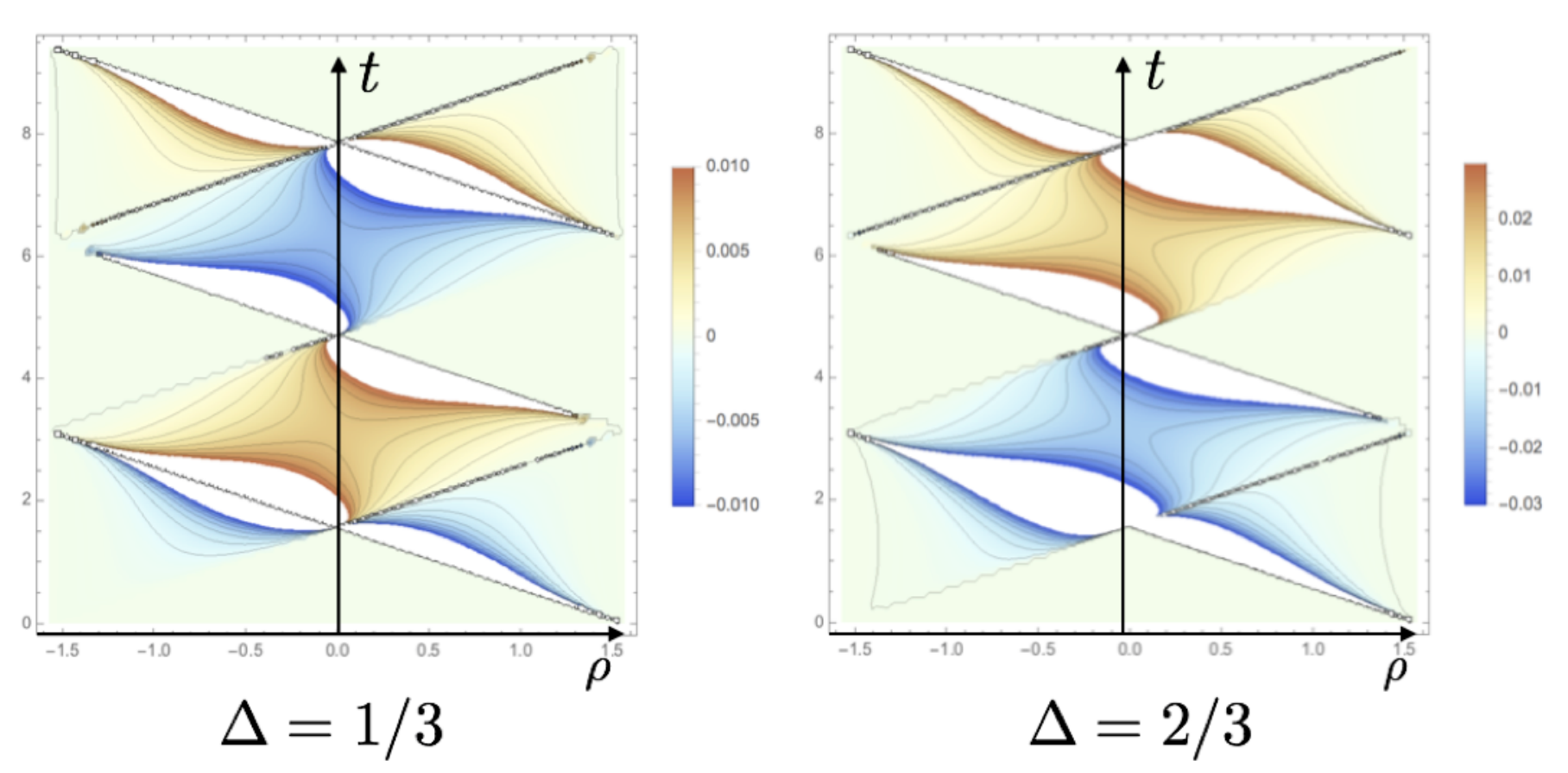} 
		        		         \label{3}
		        				 			 \caption{Plots of null energy $T_{\lambda\lambda}$ in the bulk on $\theta=0$ (and $=\pi$) plane. $U$ is taken to be $\theta$ growing direction. We can observe non zero null energy is emitted from the two points where we inserted double trace deformation. Note that on any null geodesic incoming from $\theta=\pi$ and $-\pi<\theta<0$, null energy is negative, when $g$ is positive. This result is $\Delta$ independent.}
									  \end{center} 
         \end{figure}

\begin{figure}           \begin{center}                   
		         \includegraphics[scale=0.8]{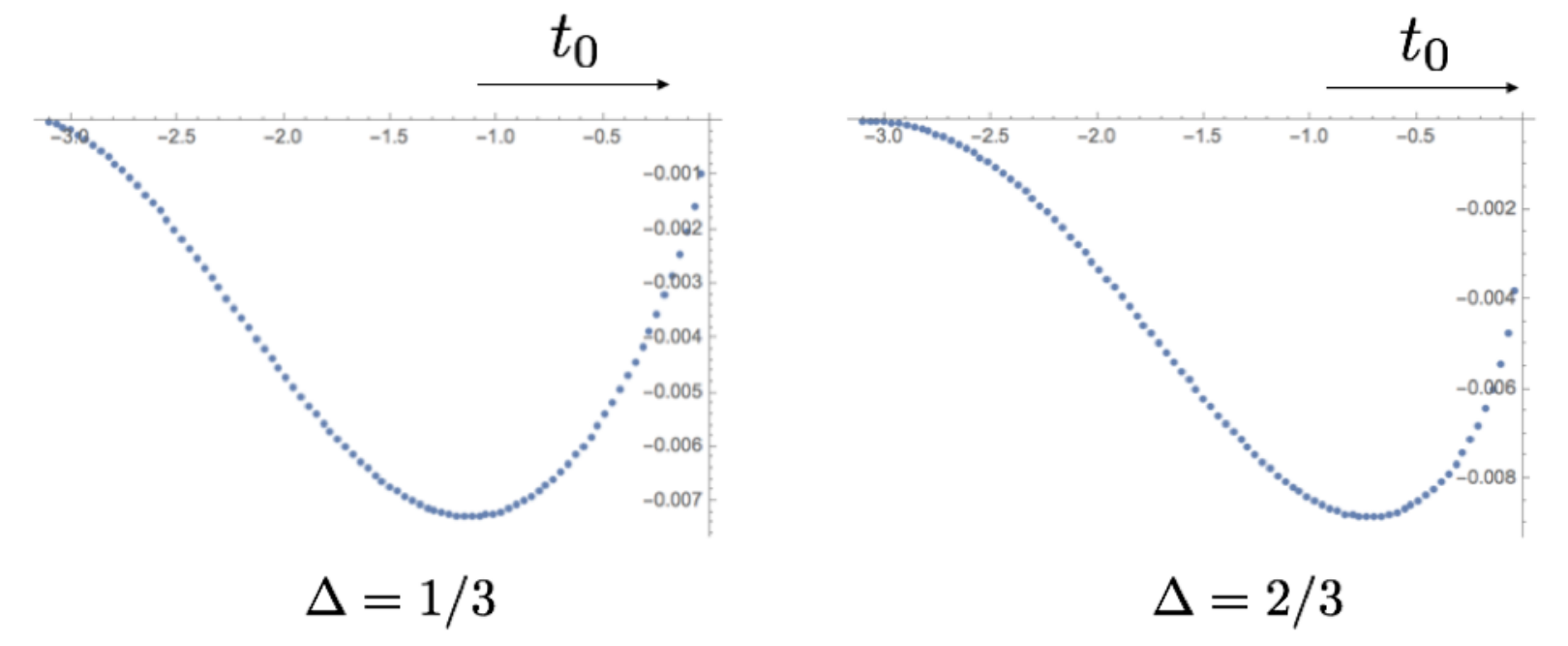} 		         
		         \label{4}
		         \end{center}
						 \caption{Plots of Averaged null energies on null geodesics when the theory is deformed by boundary double trace deformation. The null geodesic starts at boundary of AdS with $\theta=\pi$ and $t=t_0$, then ends at AdS boundary with $\theta=0$ and $t=t_o+\pi$. Averaged null energy is divergent and we regularized it using $a$ defined in appendix A. The averaged null energy is proportional to $\f{g}{a^{2\Delta}}$ and the plot is describing the coefficient of $\f{g}{a^{2\Delta}}$. Note that on any null geodesic incoming from $\theta=\pi$ and $-\pi<\theta<0$, null energy is negative, when $g$ is positive. This behavior is $\Delta$ independent.}
			 			         \end{figure}
			
\be
\rho=-{\rm Arctan}(\lambda)
,~~
t=-\rho+t_0+\f{\pi}{2}
,~~
\theta=\theta_0+\pi
\ee
for $\lambda<0$ and
\be
\rho={\rm Arctan}(\lambda)
,~~
t=\rho+t_0+\f{\pi}{2}
,~~
\theta=\theta_0
\ee
 for $\lambda>0$, where $\lambda$ is an affine parameter.
 
 Let's compute Null energy on those null geodesics
\be
\langle T_{\lambda\lambda}\rangle=\langle T_{\mu\nu}\rangle\f{dx^{\mu}}{d\lambda}\f{dx^{\nu}}{d\lambda}=
\f{\langle T_{tt}+2sign(\lambda)T_{t\rho}+T_{\rho\rho}\rangle}{(1+\lambda^2)^2}.\ee
We can compute right hand side of the equation by point splitting,
\be
\langle T_{tt}+2{\rm sign}(\lambda)T_{t\rho}+T_{\rho\rho}\rangle=\underset{x\rightarrow x'}{\rm Lim}(\partial_{\rho}\partial_{\rho'}+{\rm sign}(\lambda)(\partial_{\rho}\partial_{t'}+\partial_{\rho'}\partial_{t})+\partial_t\partial_{t'})G(x,x').
\ee
Let us take $f(t)=\delta(t)$ and $\theta_0=0$. We plot null energy $\langle T_{\lambda\lambda}\rangle$ in Figure 3 and averaged null energy $\int d\lambda\langle T_{\lambda\lambda}\rangle$ in Figure 4.         
    
We confirm that when the sign of $g$ is positive, averaged null Energy is negative for $-\pi<t_0<0$ for any $\Delta$.	So the null ray emitted into bulk in this period gets time shift and becomes time like after incorporating stress tensor in Einstein equation. Time delay in general relativity because of null energy conditions are discussed in \cite{Shapiro}\cite{Visser:1998ua}\cite{Engelhardt:2016aoo}.
Note that the averaged null energy is divergent and is proportional to $g/a^{2\Delta}$, where $a$ is lattice spacing. Naive counting of dimension of $g$ is $1-2\Delta$, so averaged null energy has naive dimension $1$.
		
 \section{Double trace deformation by bulk local operators}
 In this section, we consider multi trace deformation using HKLL bulk local operators.
 Our bulk double trace deformations model quantum operations on bulk tensor network, bulk non local interactions, and bulk micro traversable wormhole. The motivation to study this deformation is to construct wormhole inside space with single asymptotically AdS region, unlike two sided black hole with two asymptotic regions.
  Time evolution of bulk after introduction of bulk local excitations was discussed in \cite{Miyaji:2015fia}\cite{Goto:2016wme}.
   In this paper, we compute time evolution of null energy and averaged null energy, and find that both of them can be negative.
 
 \subsection{Single trace deformation}
To start with, we consider single trace deformation to Hamiltonian by HKLL bulk local operator.
Let's assume the bulk local operator sits at the center of AdS. The deformation is
\be
H\rightarrow H+g\Phi_{bulk}(t,0,0)f(t)
\ee
where $\Phi_{bulk}$ is HKLL bulk local operator.
 The vacuum expectation value of bulk local field is then given by
 \ba
\langle\Phi^{f}(t,\rho,\theta)\rangle=ig\int^t_{-\infty} d\ti{t}~f(\ti{t})\langle [\Phi(\ti{t},0,0),\Phi(t,\rho,\theta)]\rangle+\mat{O}(g^2).\ea
We assume $f(t)=\delta(t)$, then we can compute expectation value of bulk energy momentum tensors.The leading term of null energy is found to be proportional to $g^2$ with positive coefficient as shown in Figure 5. So null energy is positive everywhere and there are no violations of ANEC in this case.

\begin{figure}  
                  \begin{center}
		         \includegraphics[scale=0.95]{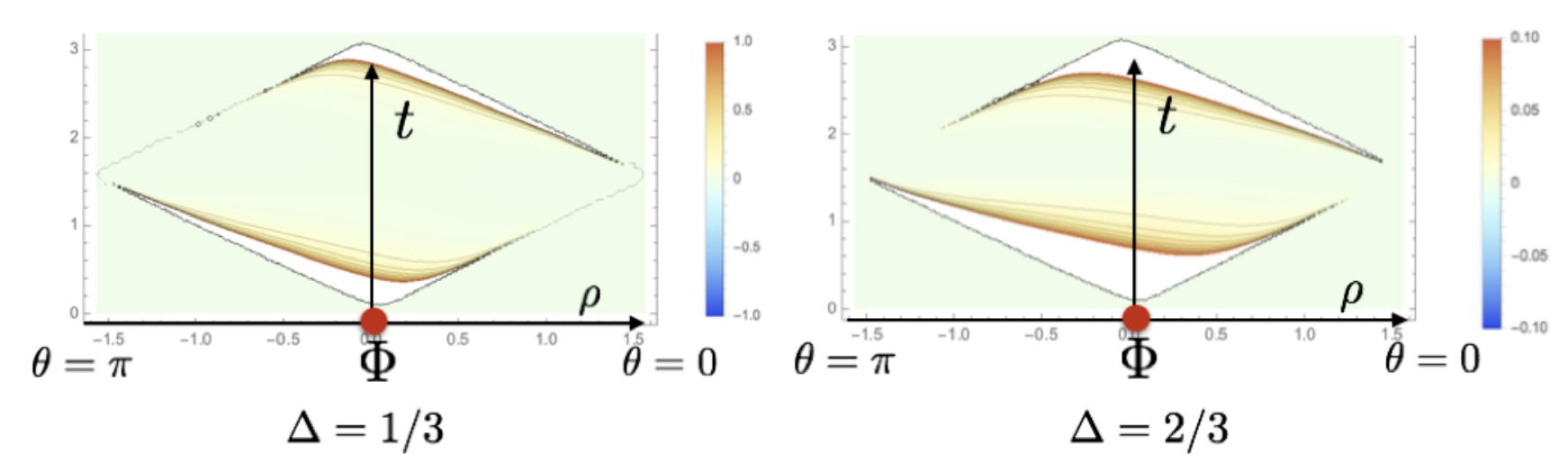} 
		         \label{5}
		        		        	 \end{center} 
			  \caption{Time evolution of null energy in the case when we introduce single trace deformation in the bulk, divided by $g^2$. We can observe positive null energy is emitted from the point where we inserted single trace deformation. In particular, null energy is always positive everywhere.}
			\end{figure}

\subsection{Double trace deformation}
In this section, we consider double trace deformation using HKLL bulk local operators, 
 in order to model traversable wormhole connecting bulk points. In this case the deformation operators live inside bulk, not on AdS boundary. 

We consider deformation of Hamiltonian
\be
H\rightarrow H+g\Phi_{bulk}(t,\rho_0,0)\Phi_{bulk}(t,\rho_0,\pi)f(t),
\ee
where $\Phi_{bulk}$ is HKLL bulk local operator.
The computation of null energy goes in the similar way as in the previous section.
The expression for perturbed bulk two point function is
\ba&&
\langle\Phi^{f}(t,\rho,\theta)\Phi^{f}(t',\rho',\theta')\rangle-\langle\Phi(t,\rho,\psi)\Phi(t',\rho',\theta')\rangle\no&&=ig\int^t_{-\infty} d\ti{t}~f(\ti{t})\langle [\Phi(\ti{t},\rho_0,0)\Phi(\ti{t},\rho_0,\pi),\Phi(t,\rho,\theta)]\Phi(t',\rho',\theta')\rangle+c.c_{(t,\rho,\theta)\leftrightarrow(t',\rho',\theta')}+\mat{O}(g^2)\no
&&\approx ig\int^t_{-\infty} d\ti{t}~f(\ti{t})\Big(\langle [\Phi(\ti{t},\rho_0,0),\Phi(t,\rho,\theta)]\rangle
\langle\Phi(\ti{t},\rho_0,\pi)\Phi(t',\rho',\theta')\rangle+(0\leftrightarrow \pi)
\Big)\no&&+c.c_{(t,\rho,\theta)\leftrightarrow(t',\rho',\theta')}+\mat{O}(g^2),\ea
which can be computed by analytical continuation of bulk two point functions\cite{Ichinose:1994rg}.

\begin{figure}     
			                         \begin{center}
		         \includegraphics[scale=1.15]{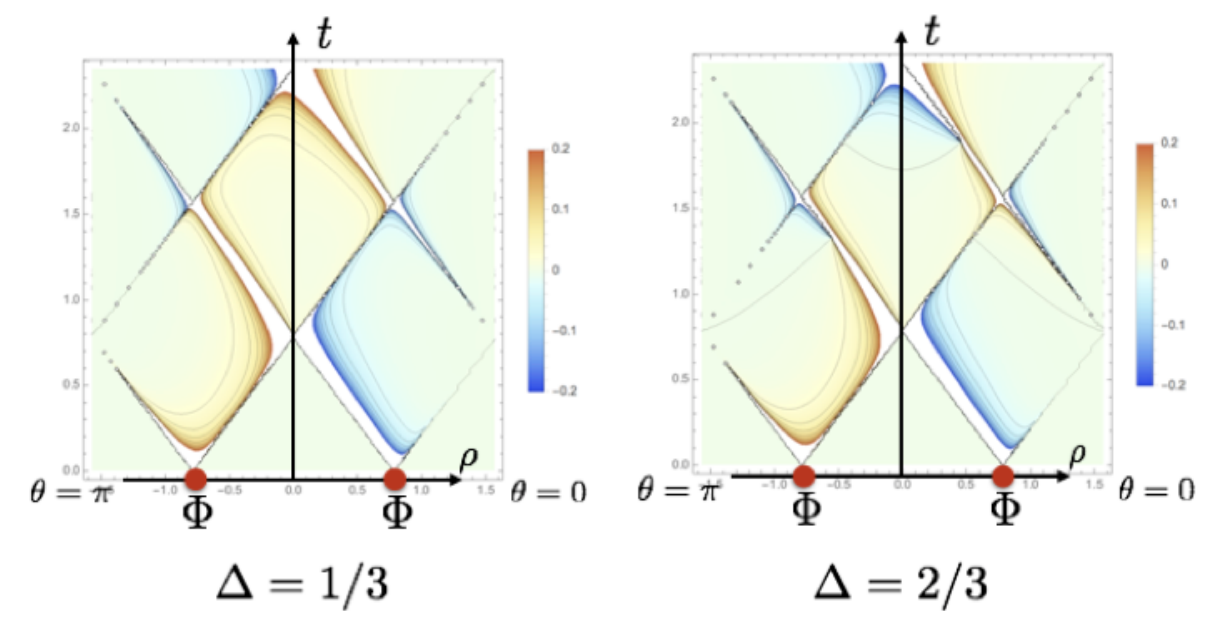} 
		        		         \label{6}		    		    
		         		        	 \end{center} 
				 \caption{Plots of null energy $T_{UU}$ in the bulk on $\theta=0$ (and $=\pi$) plane. $U$ is taken to be $\theta$ growing direction. We can observe non zero null energy is emitted from the two points where we inserted double trace deformation. Note that on any null geodesic incoming from $\theta=\pi$ and $-\pi<\theta<0$, null energy is negative, when $g$ is positive. This fact is $\Delta$ independent. The left figure is for $\Delta=1/3$ case, and the right figure is for $\Delta=2/3$ case.}
         \end{figure}
\begin{figure}  
                  \begin{center}
		         \includegraphics[scale=0.9]{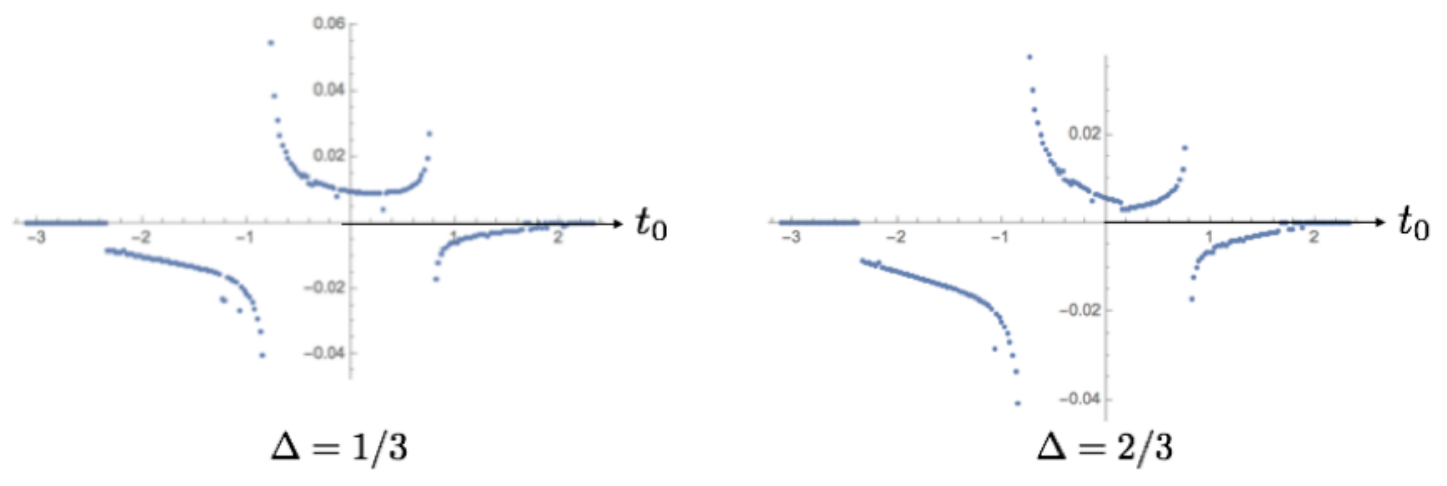} 
		          \label{7}		
		          		        		        	 \end{center} 
			  \caption{Plots of Averaged null energies on null geodesics when the theory is deformed by bulk double trace deformation. The null geodesic starts at boundary of AdS with $\theta=\pi$ and $t=t_0$, then ends at AdS boundary with $\theta=0$ and $t=t_o+\pi$. Averaged null energy is divergent and we regularized it using $a$ defined in appendix A. The averaged null energy is proportional to $\f{g}{a}$ and the plot is describing the coefficient of $\f{g}{a}$. Note that on any null geodesic incoming from $\theta=\pi$ and $-\pi<\theta<0$, null energy is negative, when $g$ is positive. This behavior is $\Delta$ independent.}
			\end{figure}

Then, by using point splitting, we can compute null energy and averaged null energy in the bulk. The results are shown in Figure 6 and Figure 7.
We confirm that when the sign of $g$ is positive, averaged null Energy is negative for $-3\pi/4<t_0<-\pi/4$ for any $\Delta$. So the null ray emitted into bulk in this period gets time shift and becomes time like after incorporating stress tensor in Einstein equation. Note that the averaged null energy is divergent and is proportional to $g/a$, where $a$ is lattice spacing. Naive counting of dimension of $g$ is $0$, so averaged null energy has naive dimension $1$.		

 One can check that when the HKLL operators approach to the boundary, then identical behavior as that of boundary local operators can be observed, as it should be.

\section{CFT analysis}
In this section, we consider time evolution of entanglement entropy and Renyi entropy in general two dimensional CFT on Minkowski space.
Entanglement entropy has been studied from various viewpoints, including from CFT\cite{Calabrese:2004eu} as well as from  holography\cite{Ryu:2006bv}. Our setup can be regarded as a version of quantum quench\cite{Calabrese:2005in}\cite{Calabrese:2007mtj}\cite{Caputa:2014vaa}\cite{Nozaki:2014uaa}\cite{Nozaki:2014hna}\cite{He:2014mwa} with non local deformation. Perturbative computation of entanglement entropy with time dependent relevant deformation is discussed in \cite{Leichenauer:2016rxw}. Entanglement entropies of non local theories are discussed in \cite{Shiba:2013jja}\cite{Mollabashi:2014qfa}.
We consider time evolution of vacuum state after acting double trace operation $e^{-ig\mat{O}(0)\mat{O}(L)}$. The non-reduced, regularized density matrix is given by
 \ba
\rho(g,t)&&=e^{-iHt}e^{-\ep H}e^{-ig\mat{O}\mat{O}}|0\rangle\langle 0|e^{ig\mat{O}\mat{O}}e^{-\ep H}e^{iHt}/\mat{N}_g
\no&&=e^{-ig\mat{O}\mat{O}(-(t-i\ep))}|0\rangle\langle 0|e^{ig\mat{O}\mat{O}(-(t+i\ep))}/\mat{N}_g,\ea
where $\mat{O}(t-i\tau)=e^{H(it+\tau)}\mat{O}e^{-H(it+\tau)}$, and $\ep$ is an infinitesimal positive constant, and $\mat{N}_g=\langle 0|e^{ig\mat{O}\mat{O}}e^{-2\ep H}e^{-ig\mat{O}\mat{O}}|0\rangle=1+\mat{O}(g^2)$. We will take subregion A as $[x,\infty]$, and consider entanglement of reduced density matrix ${\rm Tr}_{\mat{H}_{\bar{A}}}\rho(g,t)$.
 
\subsection{Entanglement Entropy}
Let us first begin with time evolution of entanglement entropy of subregion A.
Using modular Hamiltonian $H_A=-{\rm log}\rho_A$, entanglement entropy can be expressed as
\be
S_A=-{\rm Tr}[\rho_A {\rm log}\rho_A]=\langle H_{mod}^A\rangle.
\ee
In the case of vacuum modular Hamiltonian of half space in 2d CFT, it can be written explicitly in terms of energy momentum tensor\cite{Bisognano:1976za}, we have
\be
H_{mod}^A=2\pi\int_{x}^{\infty}d\ti{x}~(\ti{x}-x)T_{00}(0,\ti{x})+Const.
\ee 
Let us consider double trace deformation on CFT ground state in perturbation theory. 
Perturbation theory of entanglement entropy is discussed in \cite{Rosenhaus:2014woa}\cite{Rosenhaus:2014zza}.
From the normalization condition, we have
\be
{\rm Tr}_{\mat{H}_A}[e^{-H^A_{mod}(t,g)}\f{d}{dg}H^A_{mod}(t, g)]=0.
\ee
Then, the entanglement entropy is given by
\ba
S_A(t,g)&&={\rm Tr}_{\mat{H}_A}[e^{-H^A_{mod}(t,g)}H^A_{mod}(t, 0)]+\mat{O}(g^2)={\rm Tr}_{\mat{H}}[e^{-H_{mod}(t,g)}(H^A_{mod}(t, 0)\otimes 1_{\mat{H}_{\bar{A}}})]+\mat{O}(g^2)\no&&
=S_A(0,0)+ig\Big(\langle\mat{O}\mat{O}(-(t+i\ep))H^A(g=0)\rangle
-\langle H^A(g=0)\mat{O}\mat{O}(-(t-i\ep))\rangle\Big)+\mat{O}(g^2).\no&&
\ea
From conformal Ward-Takahashi identity, we have
\be
\langle T(z)\mat{O}(0,L)\mat{O}(0,0)\rangle=\f{2h}{L^{4h+1}}(-\f{1}{z-L}+\f{1}{z})+\f{h}{L^{4h}}(\f{1}{z^2}+\f{1}{(z-L)^2}).
\ee
In the end,
\ba&&
\Delta S_A=-2\pi ig\int_{x}^{\infty} d\ti{x} ~(\ti{x}-x)\no&&
\Big(\langle \mat{O}(-(t+i\ep),L)\mat{O}(-(t+i\ep),0)(T(0,\ti{x})+\bar{T}(0,\ti{x}))\rangle-\langle (T(0,\ti{x})+\bar{T}(0,\ti{x}))\mat{O}(-(t-i\ep),L)\mat{O}(-(t-i\ep),0)\rangle\Big).\no&&
\ea
Therefore, one need to evaluate
\be
-2\pi ig\int_{x}^{\infty} d\ti{x} ~(\ti{x}-x)\Big( \f{2h}{L^{4h+1}}(-\f{1}{z-L}+\f{1}{z})+\f{h}{L^{4h}}(\f{1}{z^2}+\f{1}{(z-L)^2})+ \f{2h}{L^{4h+1}}(-\f{1}{\bar{z}-L}+\f{1}{\bar{z}})+\f{h}{L^{4h}}(\f{1}{\bar{z}^2}+\f{1}{(\bar{z}-L)^2})\Big),
\ee
with $z=\ti{x}-(t\pm i\ep)$ and $\bar{z}=\ti{x}+(t\pm i\ep)$.
\begin{figure}
                  \begin{center}
		       \includegraphics[scale=0.6]{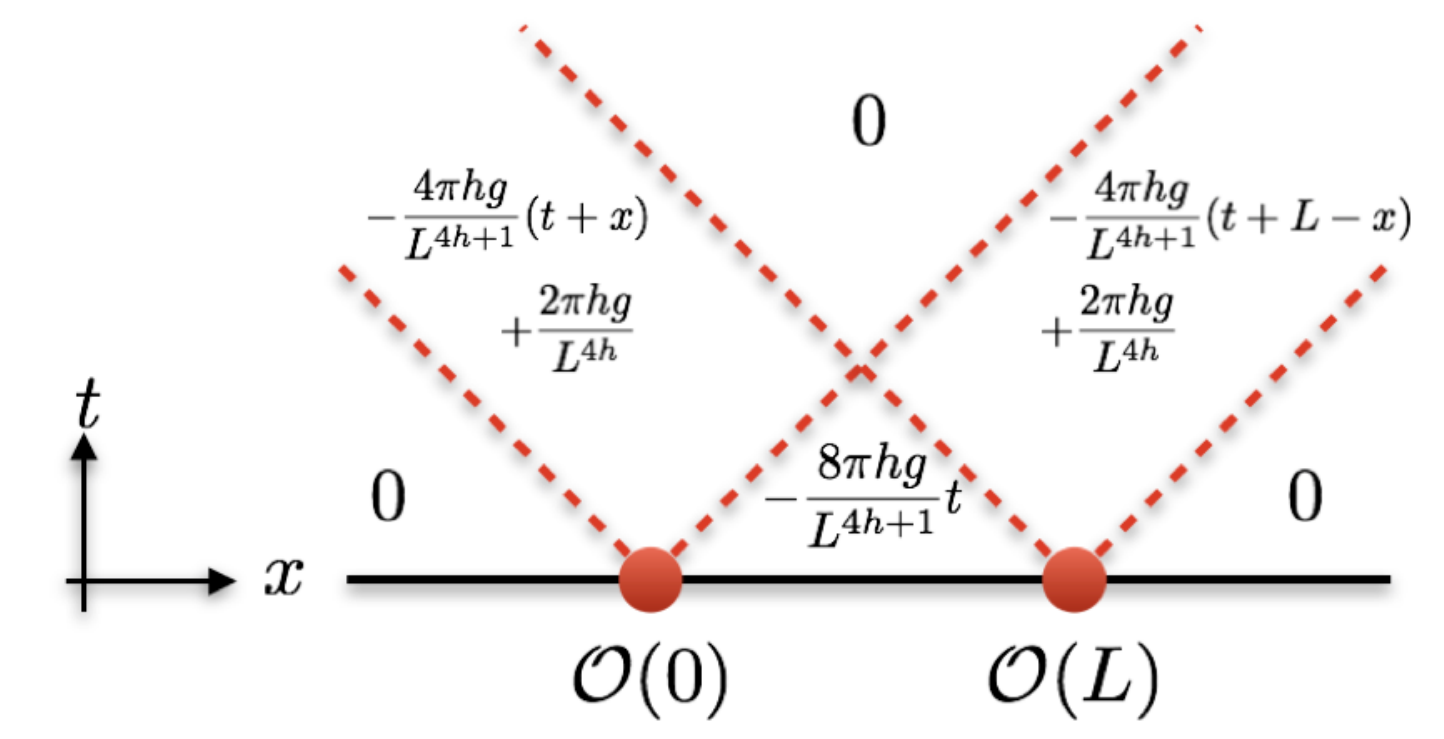} 		         
		         \label{8}
				  \end{center} 
			                			  \caption{Spacetime dependence of entanglement entropy $\Delta S_A$ after double trace deformation. When $x>t+L$ or $x+t<0$, $\Delta S_A$ must be zero from causality.}
		\end{figure}
The total shift of entanglement entropy is
\ba&&
\Delta S_A=\f{4\pi^2 gh}{L^{4h}}\Big(\theta(t-x)+\theta(t+L-x)-\theta(-t-x)-\theta(-t+L-x)\Big)\no&&
-\f{8\pi^2 gh}{L^{4h+1}}\Big((t+L-x)\theta(t+L-x)-(t-x)\theta(t-x)-(-t+L-x)\theta(-t+L-x)+(-t-x)\theta(-t-x)\Big).\no&&
\ea 
Interestingly, when $t+x-L>0$ and $t-x>0$, we have $\Delta S_A=0$. The reason why$\Delta S_A=0$ in $x-t-L>0$ and $-x+t>0$ is because of the causality. Moreover, the correction is order $g^1$, which is different from previous perturbative computation os entanglement entropy considered in \cite{Rosenhaus:2014woa}\cite{Rosenhaus:2014zza}. This is because we are using double trace and not single trace operator.

Let us compare our result with energy of half space $\int_x^\infty d\ti{x}T_{00}(t,\ti{x})$. Integral of energy density is given by
\be
\langle\int_x^\infty d\ti{x}T_{00}(t,\ti{x})\rangle=-\f{4\pi gh}{L^{4h+1}}\Big(\theta(-(x-t))-\theta(-(x-t-L))-\theta(-(x+t))+\theta(-(x+t-L))\Big).
\ee
The averaged null energy on null geodesic $(t,~x)=
\lambda, a+\lambda)$ is given by
\be
\int^{\infty}_{-\infty} d\lambda \langle T(t=\lambda,x=a+\lambda)-\ti{T}(t=\lambda,x=a+\lambda)\rangle=-\f{4\pi gh}{L^{4h+1}}\theta(L-a)\theta(a).
\ee 
When $g$ is positive, the average null energy is negative, which is consistent with the averaged null energy calculations in the bulk. Further, the perturbation is not UV divergent and is finite. Note that ANE is nonzero only when $0<a<L$. This is because of cancelation of positive energy flux and negative energy flux. Note that for excited states, the energy always increases. However, for our case, the total energy does not change at leading order since we are considering external perturbation.

\begin{figure}
 \begin{center}
		         \includegraphics[scale=0.7]{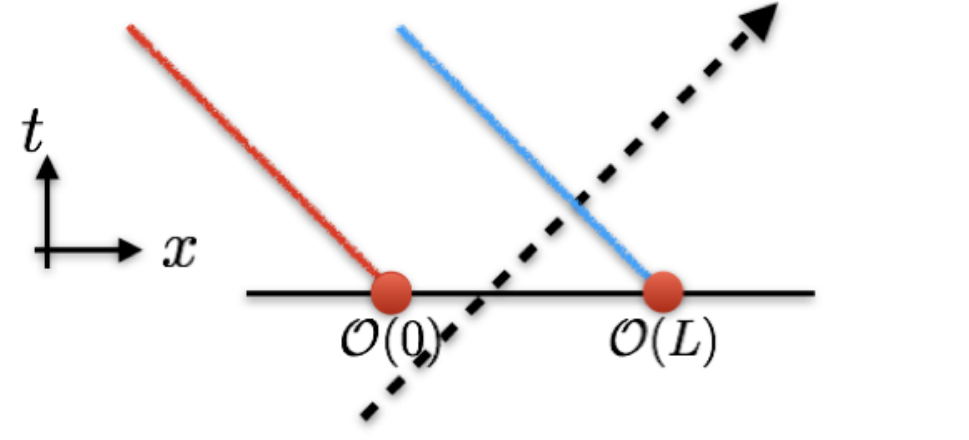} 
		        		         \label{9}
				          \caption{When null geodesic crosses with null blue line, ANE changes by $-\f{4\pi gh}{L^{4h+1}}$, and $\f{4\pi gh}{L^{4h+1}}$ with null red line. In particular, when $0<a<0$, averaged null energy is $-\f{4\pi gh}{L^{4h+1}}$, and otherwise zero.}
				       \end{center}\end{figure}				           				         
				       \begin{figure}\begin{center}
		         \includegraphics[scale=0.7]{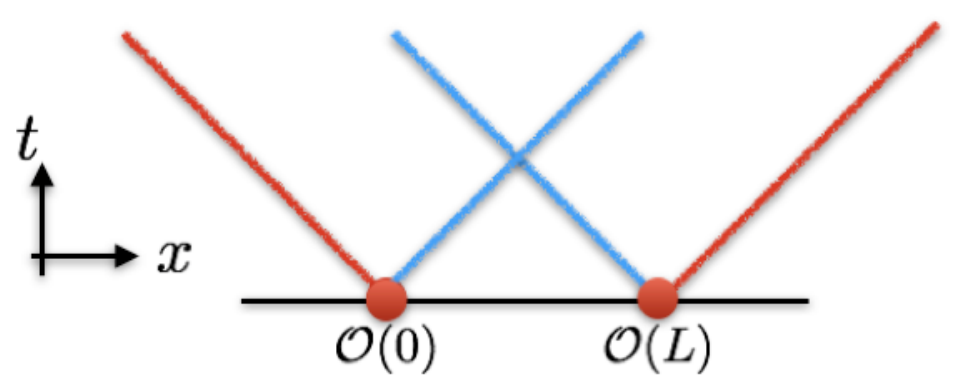} 
		        		         \label{10}
				          \caption{When constant time slice crosses null blue line, the total energy changes by $\f{4\pi gh}{L^{4h+1}}$, and $-\f{4\pi gh}{L^{4h+1}}$ with null red line. In particular, the total energy is always 0 after double trace deformation.}
				       \end{center} 
				           \end{figure}
		           \begin{figure}		
		            \begin{center}		     
				            \includegraphics[scale=1.1]{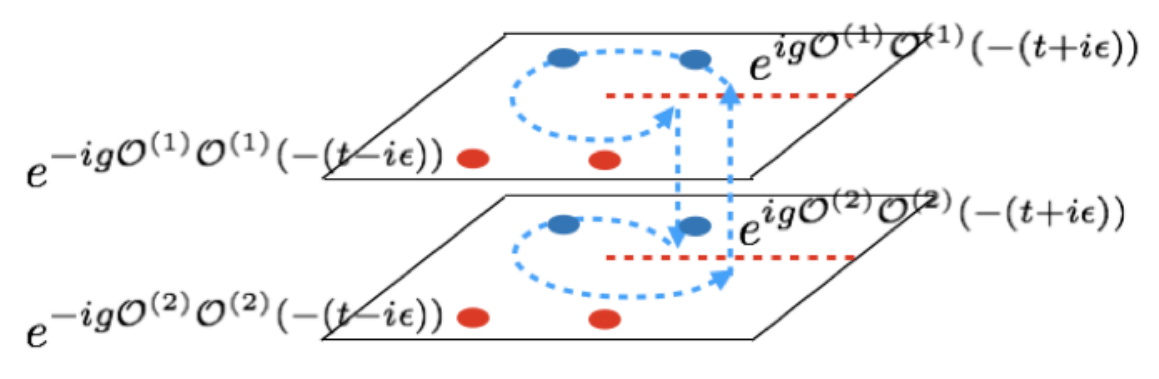} 		         
		         \label{11}
			 \caption{Geometry of replica $\Sigma_2$. Blue points correspond to $e^{ig\mat{O}\mat{O}(-(t+i\ep))}$, and red points correspond to $e^{ig\mat{O}\mat{O}(-(t-i\ep))}$.} \end{center}						  			 			         \end{figure}
												        
\subsection{Renyi Entanglement Entropy}
Next, we consider time evolution of Renyi entropy after acting double trace deformation in two dimensional CFT.
The shift of n-th Renyi entropy $\Delta S_{A}^{n}(g,t)=S_A^{n}(g,t)-S_A^n(0)$ can be computed using replica trick\cite{Calabrese:2004eu}\cite{Callan:1994py},
\be 
\Delta S_{A}^{n}(g,t)=\f{1}{1-n}
log\f{\langle e^{ig\mat{O}^{\dagger(1)}(-(t+i\ep),-x)\mat{O}^{\dagger(1)}(-(t+i\ep),-x+L)}e^{-ig\mat{O}^{(1)}(-(t-i\ep),-x)\mat{O}^{(1)}(-(t-i\ep),-x+L)}\cdots\rangle_{\Sigma_n}}
{\langle 1\rangle_{\Sigma_n}}.
\ee
where $\Sigma_n$ is n-sheeted geometry, and $\mat{O}^{(i)}$ lives on $i$-th sheet. We assume subregion as half line. Then the conformal transformation from $\Sigma_n$ to $\Sigma_1$ is
\be
w=z^n.
\ee
						       
Let us consider $g^1$ order contribution,
\ba &&\langle e^{ig\mat{O}^{\dagger(1)}\mat{O}^{\dagger(1)}(-(t+i\ep))}e^{-ig\mat{O}^{(1)}\mat{O}^{(1)}(-(t-i\ep))}\cdots\rangle_{\Sigma_n}
\no&& =\langle 1\rangle_{\Sigma_n}+ ig\sum_{i=1}^{n}\Big(\langle\mat{O}^{(i)\dagger}\mat{O}^{(i)\dagger}(-(t+i\ep))\rangle_{\Sigma_n}
-\langle\mat{O}^{(i)}\mat{O}^{(i)}(-(t-i\ep))\rangle_{\Sigma_n}\Big)+\mat{O}(g^2)\ea
Let us evaluate when $n=2$. In the first sheet, $w=-x+(t\pm i\ep)$ corresponds to
\be
z(x,\pm)=\theta(x-t)\sqrt{x-t}e^{\f{i\pi}{2}}+\theta(t-x)\sqrt{t-x}e^{\f{i\pi(1\mp 1)}{2}},
\ee
and $\bar{z}=-x-(t\pm i\ep)$ corresponds to
\be
\bar{z}(x,\pm)=\theta(x+t)\sqrt{x+t}e^{\f{-i\pi}{2}}+\theta(-t-x)\sqrt{-t-x}e^{\f{-i\pi(1\mp 1)}{2}}.
\ee
In the second sheet,  $w=-x+(t\pm i\ep)$ corresponds to
\be
z(x,\pm)=\theta(x-t)\sqrt{x-t}e^{\f{i3\pi}{2}}+\theta(t-x)\sqrt{t-x}e^{\f{i\pi(3\mp 1)}{2}},
\ee
and $\bar{z}=-x-(t\pm i\ep)$ corresponds to
\be
\bar{z}(x,\pm)=\theta(x+t)\sqrt{x+t}e^{\f{-i3\pi}{2}}+\theta(-t-x)\sqrt{-t-x}e^{\f{-i\pi(3\mp 1)}{2}}.
\ee
Then we have
\ba&&
\langle\mat{O}^{(1)}(-(t\pm i\ep),-x)\mat{O}^{(1)}(-(t\pm i\ep),-x+L)\rangle_{\Sigma_2}\no&&
=\langle\mat{O}^{(2)}(-(t\pm i\ep),-x)\mat{O}^{(2)}(-(t\pm i\ep),-x+L)\rangle_{\Sigma_2}\no&&
=(\f{d z}{d w})^h|_{z=z(x,\pm)}
 (\f{d \bar{z}}{d \bar{w}})^h|_{\bar{z}=\bar{z}(x,\pm)}\no&&
\times(\f{d z}{d w})^h|_{z=z(x-L,\pm)}
 (\f{d \bar{z}}{d \bar{w}})^h|_{\bar{z}=\bar{z}(x-L,\pm)}\no&&
\langle\mat{O}^{(1)}(z(x,\pm), 
\bar{z}(x,\pm))
\mat{O}^{(1)}(z(x-L,\pm),\bar{z}(x-L,\pm))\rangle_{\Sigma_1}.
\ea
The results are shown in Figure 12.
We can confirm that the second Renyi entropy behaves in the same way as that of entanglement entropy. In particular, we have zero Renyi entropy and zero entanglement entropy on identical regions, including regions where any change is prohibited by causality. Moreover, positive (null) energy corresponds to the growth of entanglement, and negative (null) energy corresponds to the decrease of entanglement. This is expected behavior from second law of entanglement entropy.

\begin{figure}
                  \begin{center}
		       \includegraphics[scale=1]{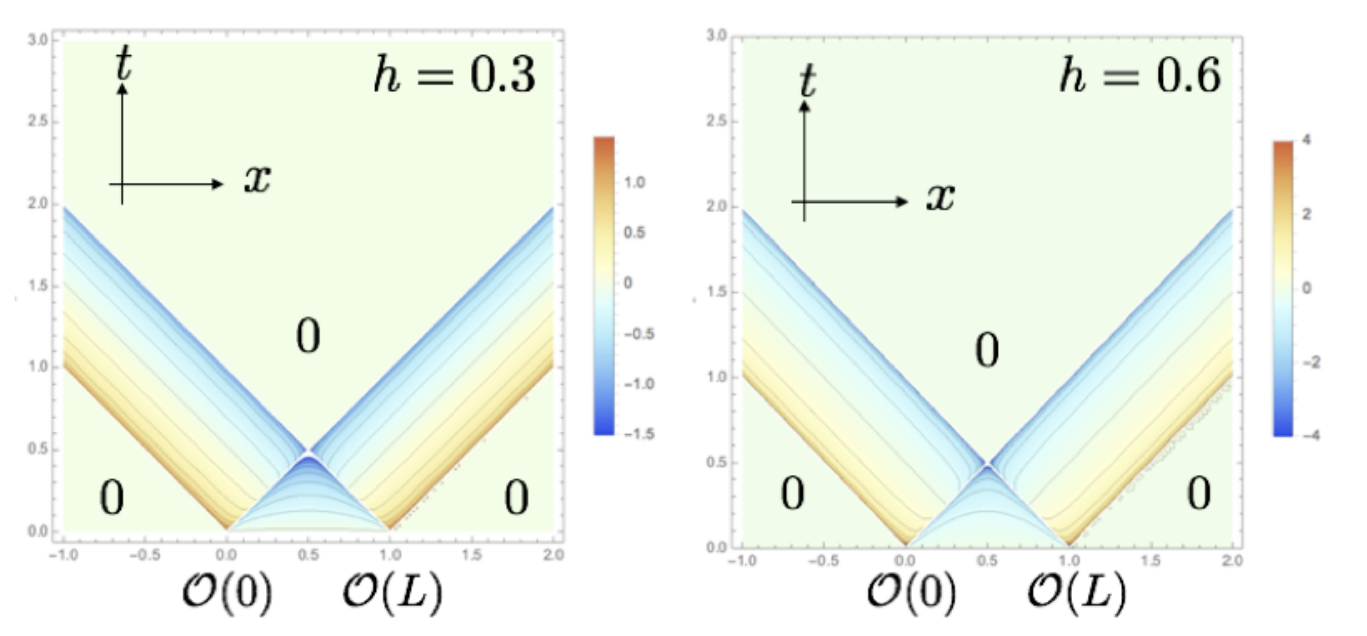} 		         
		         \label{12}
			 \caption{Entanglement Renyi entropy $\Delta S_A^2$ of semi infinite half line divided by coupling $g$. Value of Renyi entropy depends on the location of the left edge of the half line, which is specified by the coordinate $(t,x)$. When $x>t+L$ or $x+t<0$, $\Delta S_A^2$ must be zero from causality. Left figure is for $h=0.3$ and right one is for $h=0.6$. In all case one can observe that Renyi entropy behaves in similar way as that of entanglement entropy.}
			  \end{center} 
			\end{figure}

\section{Outlook}
In our study, we considered double trace deformation composed of boundary local operators as well as bulk local operators. We evaluated time shift caused by the deformation, by computing averaged null energy. Then we considered time evolution of CFT, by computing entanglement entropies and Renyi entropies.

In this paper, we only computed leading term in $g$, and entanglement introduced was order $c^0$. We can consider higher order in $g$, as well as adding more nonlocal entanglement by $e^{i\sum_lc_l\mat{O}_l}$.Generalization of our study to higher order correction might lead to the understanding of constructing wormholes in spacetime with single boundary. Also, it is interesting to apply our bulk local double trace deformation to ${\rm AdS_2}$, and see if non-pertubative 
computation is possible, although it is not clear how to define HKLL bulk local operators when bulk metric starts fluctuating.

In our computation of dynamics of entanglement entropy and entanglement Renyi entropy,
we considered instantaneous double trace deformation and leading term which is proportional to $g^1$. It is interesting and tractable to generalize our work to higher order, or to triple trace deformations. Tripartite deformation can be an interesting example to examine multi partite entanglement.

\section*{Acknowledgement}

We are grateful to Tadashi Takayanagi, Aron Wall and Pawel Caputa for useful discussions and correspondences. The author is supported by JSPS fellowship.

\appendix

\section{Propagators in AdS}
Let us summarize the propagators in global AdS${}_3$
\be
ds^2=\f{1}{cos^2\rho}(-dt^2+d\rho^2+sin^2\rho d\theta^2).
\ee
First we define 
\be
z=\f{cos(t-t')-sin\rho sin\rho' cos(\theta-\theta')}{cos\rho cos\rho'}\ee

The bulk two point function scalar of mass $m^2=\Delta(\Delta-2)$ for space-like configuration is
\ba
\langle\Phi(t,\rho,\theta)\Phi(t',\rho',\theta')\rangle=\f{1}{4\pi}(z^2-1)^{-\f{1}{2}}(z+(z^2-1)^{\f{1}{2}})^{1-\Delta}
\ea
In order to continue to time-like configulation, we need a shift $(\ep>0)$
\be
t\rightarrow t-i\ep
\ee
since we are dealing with Wightman function, not time-ordered propagator which requires $t\rightarrow t(1-i\ep)$.
In Wightman case we have 
\be
z\rightarrow z+i~\f{sin(t-t')}{cos\rho cos\rho'}\ep.
\ee

So the full expression of Wightman function is
\ba&&
\langle\Phi(t,\rho,\theta)\Phi(t',\rho',\theta')\rangle
=\theta(z-1)\f{e^{-\f{in\pi}{2}}}{4\pi}(z^2-1)^{-\f{1}{2}}(z+e^{\f{in\pi}{2}}(z^2-1)^{\f{1}{2}})^{1-\Delta}\no
\ea
where $|n|$ is the number of intersecting points, between geodesic connecting two points and light cone of $(t',\rho',\phi')$. 
And $n$ is positive when $t>t'$, and is negative when $t<t'$.

Then the commutator is
\ba&&
\langle[\Phi(t,\rho,\phi),~\Phi(t',\rho',\phi')]\rangle=2i~{\rm Im}\langle\Phi(t,\rho,\theta)\Phi(t',\rho',\theta')\rangle
\ea
The bulk-boundary two point function at spacelike separation is
\ba&&
\langle\Phi(t,\rho,\phi)\mat{O}(t',\phi')\rangle\no&&=\underset{\rho'\rightarrow \f{\pi}{2}}{Lim}\f{1}{cos^{\Delta}\rho'}\langle\Phi(t,\rho,\phi)\Phi(t',\rho',\phi')\rangle\no&&
=\f{1}{2\pi}\Big(\f{(cos\rho)/2}{cos(t-t')-sin\rho ~cos(\theta-\theta')+i~{\rm sin(t-t')}\epsilon}\Big)^{\Delta}
\ea
So the bulk-boundary two point function is
\ba&&
\langle\Phi(t,\rho,\phi)\mat{O}(t',\phi')\rangle\no&&=\f{1}{2\pi}\Big(|\f{(cos\rho)/2}{cos(t-t')-sin\rho ~cos(\theta-\theta')}|\Big)^{\Delta}
e^{-in\Delta \pi}.
\ea
$|n|$ is the number of intersecting points, between geodesic connecting two points and light cone of $(t',\phi')$. 
And $n$ is positive when $t>t'$, and is negative when $t<t'$.
Therefore the commutator is
\ba
\langle[\Phi(\tau,\rho,\phi),~\mat{O}(\tau',\phi')]\rangle
&&=\f{1}{2\pi}\Big(\f{cos\rho/2}{|cos(t-t')-sin\rho ~cos(\theta-\theta')|}\Big)^{\Delta}~(-2isin(n\Delta\pi))\no\ea

In the computation of averaged null energy, we use naive regularization with infinitesimal positive parameter $a$ as follows.
For bulk to boundary propagator, when the separation is spacelike, we use
 \ba&&
\langle\Phi(t,\rho,\phi)\mat{O}(t',\phi')\rangle\no&&=\f{1}{2\pi}\Big(\f{(cos\rho)/2}{cos(t-t')-sin\rho ~cos(\theta-\theta')+a^2}\Big)^{\Delta}e^{-in\Delta \pi},
\ea
and for timelike separation
 \ba&&
\langle\Phi(t,\rho,\phi)\mat{O}(t',\phi')\rangle\no&&=\f{1}{2\pi}\Big(\f{(cos\rho)/2}{-cos(t-t')+sin\rho ~cos(\theta-\theta')+a^2}\Big)^{\Delta}e^{-in\Delta \pi}.
\ea
For the bulk to bulk propagator at spacelike separation, we regularize as
\ba&&
\langle\Phi(t,\rho,\theta)\Phi(t',\rho',\theta')\rangle
=\f{e^{-in\pi/2}}{4\pi}(z^2-1+a^2)^{-\f{1}{2}}(z+e^{in\pi/2}(z^2-1+a^2)^{\f{1}{2}})^{1-\Delta},
\ea
and at timelike separation, we regularize as
\ba&&
\langle\Phi(t,\rho,\theta)\Phi(t',\rho',\theta')\rangle
=\f{e^{-in\pi/2}}{4\pi}(-z^2+1+a^2)^{-\f{1}{2}}(z+e^{in\pi/2}(-z^2+1+a^2)^{\f{1}{2}})^{1-\Delta}.
\ea




\begin{thebibliography}{99}
\baselineskip=8pt
\bibitem{Gao:2016bin} 
  P.~Gao, D.~L.~Jafferis and A.~Wall,
  JHEP {\bf 1712}, 151 (2017)
  doi:10.1007/JHEP12(2017)151
  [arXiv:1608.05687 [hep-th]].
  
\bibitem{Maldacena:2017axo} 
  J.~Maldacena, D.~Stanford and Z.~Yang,
  Fortsch.\ Phys.\  {\bf 65}, no. 5, 1700034 (2017)
  doi:10.1002/prop.201700034
  [arXiv:1704.05333 [hep-th]].
    
\bibitem{Almheiri:2018ijj} 
  A.~Almheiri, A.~Mousatov and M.~Shyani,
  arXiv:1803.04434 [hep-th].
    
\bibitem{deBoer:2018ibj} 
  J.~De Boer, S.~F.~Lokhande, E.~Verlinde, R.~Van Breukelen and K.~Papadodimas,
  arXiv:1804.10580 [hep-th].
       
\bibitem{Maldacena:2018lmt} 
  J.~Maldacena and X.~L.~Qi,
  arXiv:1804.00491 [hep-th].
    
\bibitem{Kelly:2014mra} 
  W.~R.~Kelly and A.~C.~Wall,
  Phys.\ Rev.\ D {\bf 90}, no. 10, 106003 (2014)
  Erratum: [Phys.\ Rev.\ D {\bf 91}, no. 6, 069902 (2015)]
  doi:10.1103/PhysRevD.90.106003, 10.1103/PhysRevD.91.069902
  [arXiv:1408.3566 [gr-qc]].
     
\bibitem{Faulkner:2016mzt} 
  T.~Faulkner, R.~G.~Leigh, O.~Parrikar and H.~Wang,
  JHEP {\bf 1609}, 038 (2016)
  doi:10.1007/JHEP09(2016)038
  [arXiv:1605.08072 [hep-th]].
    
\bibitem{Hartman:2016lgu} 
  T.~Hartman, S.~Kundu and A.~Tajdini,
  JHEP {\bf 1707}, 066 (2017)
  doi:10.1007/JHEP07(2017)066
  [arXiv:1610.05308 [hep-th]].
    
\bibitem{Bousso:2015wca} 
  R.~Bousso, Z.~Fisher, J.~Koeller, S.~Leichenauer and A.~C.~Wall,
  Phys.\ Rev.\ D {\bf 93}, no. 2, 024017 (2016)
  doi:10.1103/PhysRevD.93.024017
  [arXiv:1509.02542 [hep-th]].
    
\bibitem{Koeller:2015qmn} 
  J.~Koeller and S.~Leichenauer,
  Phys.\ Rev.\ D {\bf 94}, no. 2, 024026 (2016)
  doi:10.1103/PhysRevD.94.024026
  [arXiv:1512.06109 [hep-th]].
  
\bibitem{Wall:2017blw} 
  A.~C.~Wall,
  Phys.\ Rev.\ Lett.\  {\bf 118}, no. 15, 151601 (2017)
  doi:10.1103/PhysRevLett.118.151601
  [arXiv:1701.03196 [hep-th]].
                
\bibitem{Balakrishnan:2017bjg} 
  S.~Balakrishnan, T.~Faulkner, Z.~U.~Khandker and H.~Wang,
  arXiv:1706.09432 [hep-th].
   
    
  \bibitem{Hamilton:2006az} 
  A.~Hamilton, D.~N.~Kabat, G.~Lifschytz and D.~A.~Lowe,
  Phys.\ Rev.\ D {\bf 74}, 066009 (2006)
  doi:10.1103/PhysRevD.74.066009
  [hep-th/0606141].
  
\bibitem{Maldacena2018} 
  J.~Maldacena, A.~Milekhin and F.~Popov,
  arXiv:1807.04726 [hep-th].
  
\bibitem{Ichinose:1994rg} 
  I.~Ichinose and Y.~Satoh,
  Nucl.\ Phys.\ B {\bf 447}, 340 (1995)
  doi:10.1016/0550-3213(95)00197-Z
  [hep-th/9412144].

\bibitem{Shapiro} 
Shapiro, I. I. Fourth test of general relativity. Phys. Rev. Lett. 13, 789–791 (1964).

\bibitem{Visser:1998ua} 
  M.~Visser, B.~Bassett and S.~Liberati,
  Nucl.\ Phys.\ Proc.\ Suppl.\  {\bf 88}, 267 (2000)
  doi:10.1016/S0920-5632(00)00782-9
  [gr-qc/9810026].
  
\bibitem{Engelhardt:2016aoo} 
  N.~Engelhardt and S.~Fischetti,
  Class.\ Quant.\ Grav.\  {\bf 33}, no. 17, 175004 (2016)
  doi:10.1088/0264-9381/33/17/175004
  [arXiv:1604.03944 [hep-th]].
    
\bibitem{Miyaji:2015fia} 
  M.~Miyaji, T.~Numasawa, N.~Shiba, T.~Takayanagi and K.~Watanabe,
  Phys.\ Rev.\ Lett.\  {\bf 115}, no. 17, 171602 (2015)
  doi:10.1103/PhysRevLett.115.171602
  [arXiv:1506.01353 [hep-th]].
  
\bibitem{Goto:2016wme} 
  K.~Goto, M.~Miyaji and T.~Takayanagi,
  JHEP {\bf 1609}, 130 (2016)
  doi:10.1007/JHEP09(2016)130
  [arXiv:1605.02835 [hep-th]].
    
\bibitem{Calabrese:2004eu} 
  P.~Calabrese and J.~L.~Cardy,
  J.\ Stat.\ Mech.\  {\bf 0406}, P06002 (2004)
  doi:10.1088/1742-5468/2004/06/P06002
  [hep-th/0405152].
     
\bibitem{Ryu:2006bv} 
  S.~Ryu and T.~Takayanagi,
  Phys.\ Rev.\ Lett.\  {\bf 96}, 181602 (2006)
  doi:10.1103/PhysRevLett.96.181602
  [hep-th/0603001].

\bibitem{Calabrese:2005in} 
  P.~Calabrese and J.~L.~Cardy,
  J.\ Stat.\ Mech.\  {\bf 0504}, P04010 (2005)
  doi:10.1088/1742-5468/2005/04/P04010
  [cond-mat/0503393].
  
\bibitem{Calabrese:2007mtj} 
  P.~Calabrese and J.~Cardy,
  J.\ Stat.\ Mech.\  {\bf 0710}, no. 10, P10004 (2007)
  doi:10.1088/1742-5468/2007/10/P10004
  [arXiv:0708.3750 [quant-ph]].
  
\bibitem{Caputa:2014vaa} 
  P.~Caputa, M.~Nozaki and T.~Takayanagi,
  PTEP {\bf 2014}, 093B06 (2014)
  doi:10.1093/ptep/ptu122
  [arXiv:1405.5946 [hep-th]].
  
\bibitem{Nozaki:2014uaa} 
  M.~Nozaki,
  JHEP {\bf 1410}, 147 (2014)
  doi:10.1007/JHEP10(2014)147
  [arXiv:1405.5875 [hep-th]].
  
      
\bibitem{Nozaki:2014hna} 
  M.~Nozaki, T.~Numasawa and T.~Takayanagi,
  Phys.\ Rev.\ Lett.\  {\bf 112}, 111602 (2014)
  doi:10.1103/PhysRevLett.112.111602
  [arXiv:1401.0539 [hep-th]].
  
\bibitem{He:2014mwa} 
  S.~He, T.~Numasawa, T.~Takayanagi and K.~Watanabe,
  Phys.\ Rev.\ D {\bf 90}, no. 4, 041701 (2014)
  doi:10.1103/PhysRevD.90.041701
  [arXiv:1403.0702 [hep-th]].
  
\bibitem{Leichenauer:2016rxw} 
  S.~Leichenauer, M.~Moosa and M.~Smolkin,
  JHEP {\bf 1609}, 035 (2016)
  doi:10.1007/JHEP09(2016)035
  [arXiv:1604.00388 [hep-th]].
    
\bibitem{Shiba:2013jja} 
  N.~Shiba and T.~Takayanagi,
  JHEP {\bf 1402}, 033 (2014)
  doi:10.1007/JHEP02(2014)033
  [arXiv:1311.1643 [hep-th]].
  
\bibitem{Mollabashi:2014qfa} 
  A.~Mollabashi, N.~Shiba and T.~Takayanagi,
  JHEP {\bf 1404}, 185 (2014)
  doi:10.1007/JHEP04(2014)185
  [arXiv:1403.1393 [hep-th]].
 
\bibitem{Bisognano:1976za} 
  J.~J.~Bisognano and E.~H.~Wichmann,
  J.\ Math.\ Phys.\  {\bf 17}, 303 (1976).
  doi:10.1063/1.522898

\bibitem{Rosenhaus:2014woa} 
  V.~Rosenhaus and M.~Smolkin,
  JHEP {\bf 1412}, 179 (2014)
  doi:10.1007/JHEP12(2014)179
  [arXiv:1403.3733 [hep-th]].
      
\bibitem{Rosenhaus:2014zza} 
  V.~Rosenhaus and M.~Smolkin,
  JHEP {\bf 1502}, 015 (2015)
  doi:10.1007/JHEP02(2015)015
  [arXiv:1410.6530 [hep-th]].
  
\bibitem{Callan:1994py} 
  C.~G.~Callan, Jr. and F.~Wilczek,
  Phys.\ Lett.\ B {\bf 333}, 55 (1994)
  doi:10.1016/0370-2693(94)91007-3
  [hep-th/9401072].
    
    
    
    
    
    


                              \end{thebibliography}
\end{document}